\documentclass[letterpaper,twocolumn,10pt]{article}
\usepackage[table,xcdraw]{xcolor}
\usepackage{usenix2019_v3}

\usepackage{todonotes}
\usepackage{endnotes}
\usepackage{url}
\usepackage{listings}
\usepackage{enumitem}
\usepackage[frozencache,cachedir=minted-cache]{minted}
\usepackage{mdframed}
\usepackage{multirow}
\usepackage[nolist,nohyperlinks]{acronym} 
\usepackage{xspace}
\usepackage{amsmath}
\usepackage{amssymb}
\usepackage[compatibility=false]{caption}
\usepackage{changepage}

\usepackage{afterpage}

\hypersetup{draft}

\begin{document}

\date{}

\newcounter{cs}
\newcounter{cj}
\newcounter{jc}

\newcommand{\cs}[1]{\textcolor{red}{\{CS-\arabic{cs}: #1\}}\addtocounter{cs}{1}}
\newcommand{\chani}[1]{\textcolor{red}{\{Chani-\arabic{cj}: #1\}}\addtocounter{cj}{1}}
\newcommand{\jake}[1]{\textcolor{red}{\{Jake-\arabic{jc}: #1\}}\addtocounter{jc}{1}}

\newcommand{\methodname}{Token-Level Fuzzing}
\newcommand{\toolname}{\xspace{Token-Level AFL}}
\newcommand{\numbugs}{\xspace{29}}

\newcommand{\ignore}[1]{}

%
\title{\Large \bf Token-Level Fuzzing}


\author{
{\rm Christopher Salls}\\
UC Santa Barbara
\and
{\rm Chani Jindal}\\
Microsoft
\and
{\rm Jake Corina}\\
Seaside Security
\and
{\rm Christopher Kruegel}\\
UC Santa Barbara
\and
{\rm Giovanni Vigna}\\
UC Santa Barbara
} 


\maketitle

\begin{abstract}
Fuzzing has become a commonly used approach to identifying bugs in complex, real-world programs. 
However, \emph{interpreters} are notoriously difficult to fuzz effectively, as they expect highly structured inputs, which are rarely produced by most fuzzing mutations.
For this class of programs, grammar-based fuzzing has been shown to be effective.
Tools based on this approach can find bugs in the code that is executed after parsing the interpreter inputs, by following language-specific rules when generating and mutating test cases.

Unfortunately, grammar-based fuzzing is often unable to discover subtle bugs associated with the parsing and handling of the language syntax.
Additionally, if the grammar provided to the fuzzer is incomplete, or does not match the implementation completely, the fuzzer will fail to exercise important parts of the available functionality.

In this paper, we propose a new fuzzing technique, called \methodname{}. 
Instead of applying mutations either at the byte level or at the grammar level, \methodname{} applies mutations at the \emph{token level}.
Evolutionary fuzzers can leverage this technique to both generate inputs that are parsed successfully \emph{and} generate inputs that do not conform strictly to the grammar.
As a result, the proposed approach can find bugs that neither byte-level fuzzing nor grammar-based fuzzing can find.
We evaluated \methodname{} by modifying AFL and fuzzing four popular JavaScript engines, finding \numbugs{} previously unknown bugs, several of which could not be found with state-of-the-art byte-level and grammar-based fuzzers.

\end{abstract}

\section{Introduction}

As the amount of software in the world grows, so does the need for effective bug-finding techniques.
Unfortunately, it is very common for companies to employ far more developers than security engineers.
BSIMM, a study of software security initiatives started by Synopsys, found that there was an average ratio of a single security engineer for every sixty software developers~\cite{bsimm8}.
Consequently, security engineers are often responsible for very large amounts of code; far more than is feasible to check manually.
As a result, it is imperative that effective automated techniques are used to identify security bugs.

In the past few years, fuzz testing has become widely popular.
Fuzzers such as American Fuzzy Lop (AFL)~\cite{afl}, Syzkaller~\cite{syzkaller}, and Libfuzzer~\cite{libfuzzergv} are responsible for the detection of hundreds of high-severity security issues.
The success of these fuzzers, as well as others, has caused fuzzing to become a preeminent automated analysis for detecting memory corruption vulnerabilities.
Fuzzing is employed by companies and organizations both for finding old bugs and as an additional test in continuous integration systems~\cite{continuousfuzzing}. 

The popularity of fuzzing has inspired a vast amount of research to develop new techniques, tailored to a variety of targets.
A particularly interesting target is represented by \emph{interpreters}.
Interpreters are in widespread use; they are found in many components of browsers, document viewers, programming languages, and more.
As such, interpreters are often a high-value target for attackers, and a high-impact topic for security researchers.

Analyzing interpreters is challenging: Modern-day interpreters can be very complex (for example, V8, Google's JavaScript engine, contains over 700K lines of code), and, in addition,  they expect highly structured inputs, composed of individual \emph{tokens}.
If the input does not match the format that the interpreter is expecting, it may throw an error very early in the processing (input parsing) step.
As a result, many of the most common fuzzers fail to perform well when applied to interpreters, such as JavaScript engines, because their  mutations typically result in simple syntax errors.

Because of the aforementioned issue, some fuzzers, which are targeted at the analysis of interpreters, use grammar-based approaches to generate and mutate inputs~\cite{godefroid2008grammar, wang2019superion, fuzzingjsfunpwnage, guo2017mongodb}. Their goal is to generate inputs that exercise deeper code paths. 
While these approaches are effective, they also suffer from important limitations.
First, they need to be given or be able to learn a grammar, which makes it difficult to re-target them for different languages.
Another limitation is that grammar-based fuzzers frequently conform too tightly to the supplied grammar and fail to generate unusual situations for the parser, potentially missing subtle bugs related to the syntactic analysis. 

In this paper, we introduce a novel technique, called \emph{\methodname{}}.
\methodname{} can be thought of as a level in between the byte-level approaches and the grammar-based approaches typically employed by fuzzers.
The basic idea behind \methodname{} is to have the mutations operate with whole tokens, either replacing or inserting entire words.
For example, instead of replacing a few random bytes, which has a small chance of producing an interesting input, a token-level approach would replace a few tokens in the input with different tokens, without taking into account grammar rules.
This approach allows the fuzzer to have a much higher chance of producing useful mutations, while avoiding the strictness and complexity of grammar-based approaches.

We created a modified version of AFL, called \emph{\toolname{}}, which implements \methodname{}.
Even though \toolname{} is specifically implemented for fuzzing JavaScript interpreters, the \methodname{} technique itself is general.
We tested \toolname{} against the most up-to-date versions of four major JavaScript interpreters, namely, V8, SpiderMonkey, JavaScriptCore, and ChakraCore, and we discovered \numbugs{} previously unknown bugs, some of which are severe and can lead to remote code execution.

In summary, we make the following contributions:

\begin{itemize}
	\item We introduce a new technique, called \methodname{}, for fuzzing language-based programs, such as interpreters;
	\item We implemented this technique to fuzz JavaScript engines. The implementation is done on top of AFL to take advantage of its efficient coverage-guided fuzzing;
	\item We evaluated \toolname{} on the latest versions of four major JavaScript engines, finding \numbugs{} previously unknown bugs;
	\item We compared our tool to other state-of-the-art JavaScript fuzzers, demonstrating that our tool is more effective at finding bugs.
\end{itemize}




\section{Background and Related Work}
\label{sec:relatedwork}

Fuzzing is one of the most effective and scalable vulnerability discovery solutions. 
Fuzzers generate a vast number of test cases to exercise target applications and monitor their run-time execution to discover security bugs. 
Most fuzzing research can be characterized across three axes: Input generation, program access, and coverage goals.

\textbf{Input Generation.} There are two main classes of approaches to generate inputs: mutational fuzzing and generational fuzzing.
Mutational fuzzing~\cite{gross2018fuzzil, rawat2017vuzzer, cha2015program} modifies seeds of typically well-formed inputs to generate new inputs.
Generational fuzzing, on the other hand, tends to be more structure-aware and leverages descriptions of the input format to generate inputs following that structure~\cite{han2017imf, han2019codealchemist, grieco2016quickfuzz}.

\textbf{Program Access.} Fuzzing approaches might differ in the level of insight they have into the execution of a target program.
White-box fuzzing performs program analyses and collects constraints from conditional branches during execution. 
Solutions obtained from solving these constraints are then mapped to new inputs~\cite{godefroid2012sage, stephens2016driller}.
Black-box fuzzing approaches, instead, do not have any access to the internals of the program being tested~\cite{doupe12:enemy-of-the-state, wang2019discovering}.
Finally, in the middle, are grey-box fuzzing approaches, which use lightweight techniques to gather information about program execution, such as branch coverage~\cite{fuzzingbook}.

\textbf{Coverage Goals.} Fuzzing approaches might have different goals when exercising a target program.
For example, directed fuzzing has the objective of targeting a set of deep paths~\cite{bohme2017directed}.
Coverage-based fuzzing, instead, uses different types of tracking such as block coverage, edge coverage, etc., to track the inputs that maximize code coverage, so that they are used as a basis for further mutations~\cite{afl, syzkaller, chen2018angora}.

\medskip

In the following, we provide more details about a subclass of mutational fuzzing approaches, called \emph{evolutionary} approaches, and how they are applied to JavaScript fuzzing, as JavaScript interpreters are the target of our prototype.

\subsection{Evolutionary Fuzzing}

American Fuzzy Lop (AFL) is a grey-box fuzzer that leverages compile-time instrumentation~\cite{afl} to collect meta-information about a target program's execution. 
AFL has been demonstrated to be extremely effective in finding vulnerabilities and other interesting bugs in many applications~\cite{aflbugs}. 
The main insight behind AFL's success is that inputs that exercise new paths in a program are best suited for fostering the bug-discovery process. 
Therefore, whenever AFL identifies an input that discovers a new path in the program, it uses that input as a basis for additional mutations, to see if these ``evolved'' inputs can cause the program to execute additional portions of code (measured in basic blocks). 

This evolutionary approach has been extremely effective, and, therefore, there has been much research work on improving evolutionary fuzzing. 
For example, Vuzzer~\cite{rawat2017vuzzer} focuses on extracting two main features, namely data-flow features (using taint analysis) and control-flow features, to create a smart feedback loop. 
These features, which are extracted using static analysis, help infer important properties of the inputs and prioritize/de-prioritize certain paths. 
AFLFAST, on the other hand, uses a Markov-chain-based search strategy to choose low-frequency paths, enabling the tool to explore more paths in the same fuzzing time~\cite{bohme2017coverage}. 
Another evolutionary approach is Angora~\cite{chen2018angora}, which uses byte-level taint tracking and gradient-based search algorithms in addition to input length exploration and context-sensitive branch count.

\subsection {JavaScript Fuzzing}
JavaScript engines are one of the most complicated components of modern-day browsers, making them a very popular target for both attackers and researchers. 
As a consequence, there have been continuous efforts towards improving fuzzing approaches to find JavaScript engine vulnerabilities. 
Most of these approaches fall into two categories: Grammar-based approaches and coverage-guided approaches.

\subsubsection{Grammar-Based Approaches}

Some of the most popular JavaScript fuzzers have been centered around generating syntactically correct test cases based on either a predefined grammar or a trained probabilistic language model. 
JSFunFuzz is one such JavaScript grammar-based fuzzers. 
JSFunFuzz relies purely on a generative approach to create new test cases~\cite{jsfunfuzz}, and has been used to exercise a wide range of JavaScript language features. 
Another example of a generative approach is Domato~\cite{fratric2017great}, which uses HTML, CSS, and JavaScript grammars to generate samples that target DOM-specific logic issues. 

A different approach is followed by CodeAlchemist~\cite{han2019codealchemist}, which uses semantics-aware assembly to produce JavaScript code snippets. 
This approach breaks JavaScript seeds into code fragments, and each fragment is tagged with constraints and analyzed for used variables. 
The code fragments are then combined to produce syntactically and semantically correct test cases. 

LangFuzz~\cite{holler2012fuzzing} also employed the concept of code fragments, combined with both generative and mutation-based fuzzing, to maintain the syntax and semantics of code samples. 
One key feature of LangFuzz is that it is language-independent, which means that it bases its testing strategy solely on grammar and existing programs and not language-specific information. 

\subsubsection{Coverage-Guided Approaches}

Coverage-guided fuzzing has also been successful in finding JavaScript engine vulnerabilities. 
In these approaches, one of the most common targets for mutation is the Abstract Syntax Tree (AST) of JavaScript programs~\cite{guo2017mongodb, wang2019superion}. 
For example, Fuzzilli~\cite{gross2018fuzzil} developed an intermediate language, called FuzzIL, which supports better control-flow-based and data-flow-based decisions in the mutation process. 

Nautilus~\cite{aschermann2019nautilus} is another tool that performs mutations on the ASTs, with its unique point being that it also performs byte mutations on the raw code strings.
Montage~\cite{lee2020montage} also leverages the idea that fragments of the ASTs from the test cases can be combined in unique ways to find bugs, and they do so using machine learning.
Superion~\cite{wang2019superion} modifies AFL to perform fuzzing on the ASTs, using custom grammar-based mutation strategies to achieve both grammar-aware trimming and tree-based mutations.
We compare our approach against Superion in Section~\ref{sec:comparison}, showing that our approach finds more bugs and has better code coverage.

Another work on JavaScript fuzzing that operates on ASTs was presented by Park, et al.~\cite{park2020fuzzing}.
The authors introduce the concept of aspect-preserving mutations.
Their fuzzer, called \emph{Die}, centers around the idea that there are key properties, or \emph{aspects}, in the seeds present in test cases or other bug reports.
The goal of this technique is to keep these beneficial properties from the original seed and retain them across mutations. 
For example, control-flow structures, like loops, can trigger JIT compilation, which, in turn, might reveal a buggy optimization logic; therefore, control-flow structures are an aspect that the fuzzer should specifically try to preserve during the mutation process.
This fuzzer also performs mutations in a grammar-aware manner, with mutations performed on the ASTs.

\section{Motivation}
\label{sec:motivation}




As discussed in the previous section, fuzzing research has come quite a long way from just generating purely random input.
AFL, in particular, is a venerable fuzzer that has been instrumental in finding many bugs in over one hundred highly used targets.
However, when AFL is applied to interpreters, such as JavaScript engines, some significant downsides begin to emerge.
As most of the mutations that AFL performs are at a byte- or bit-level, we see it repeatedly generating inputs that simply fail to parse.


If we consider a simple bit-flip mutation on a small piece of JavaScript, the results will frequently look like the following mutations, which will immediately fail to parse:

\begin{lstlisting}[escapeinside={(*}{*)}]
while (bar.x) (*$\longrightarrow$*) whkle (bar.x)
              (*$\longrightarrow$*) whilep(bar.x)
              (*$\longrightarrow$*) while xbar.x)
              (*$\longrightarrow$*) while (bar.|)       
\end{lstlisting}

It should be straightforward to see that mutations such as these are not particularly helpful; they will only cause simple syntax errors.
As such, this approach to mutating the inputs would very likely not lead to more code coverage, and would simply waste execution time.



The ineffectiveness of byte-level fuzzing suggests that taking into account the rules governing input format might allow for a more comprehensive exploration of a JavaScript interpreter's code base. 
Grammar-based fuzzers are incredibly powerful in their ability to very quickly generate syntactically correct pieces of input for a given program.
An obvious downside with this approach, however, is the work required to first define a grammar, or otherwise rely on an existing grammar definition before fuzzing can be performed~\cite{han2019codealchemist, wang2019superion,holler2012fuzzing,peachfuzz,boofuzz}.

\begin{figure}[ht]
  \centering
\begin{minted}{js}
function main() {
  const v1 = [13.37,13.37,13.37];
  const v6 = [1337,1337,v1];  
  function v9(v10,...v1) {
    const v13 = [1337,1337,1337];
    return v13;
  }  
  const v18 = v9(v6);
}
main();
\end{minted}
  \captionof{listing}{Example of code generated by Fuzzilli. Fuzzilli follows a static single-assignment format for the generated code. As such, variables will always be assigned exactly once and some syntactic/semantic patterns cannot be emitted.}
  \label{lst:fuzzilli}
\end{figure}

An additional downside to grammar-based fuzzing is the adherence to the grammar that is given to the fuzzer.
This not only limits the fuzzer to creating code that matches the grammar, but it \emph{also} limits the fuzzer to finding bugs that can be expressed as such.
This will prevent most grammar-based fuzzers from finding bugs that \emph{require} syntactically or semantically incorrect inputs to be triggered.
Even bugs with unusual semantics can be unreachable by grammar-based fuzzers.
This is because a grammar-based fuzzer, though powerful in its generational capabilities and language awareness, will only generate inputs that adhere to the grammar that has been supplied.

To explain this limitation further, we will show an example from Fuzzilli and talk about how its grammar limits the bugs it can find.
Listing~\ref{lst:fuzzilli} shows an example input generated by Fuzzilli, which was taken when fuzzing a JavaScript engine.
Note how each line assigns at most a single new variable, and variables are never overwritten.
This is because Fuzzilli uses a static single-assignment intermediate representation~\cite{gross2018fuzzil}, and the inputs it generates will conform tightly to it.
While this feature is instrumental in achieving the the real-world results that Fuzzilli has published, it also limits the sorts of bugs that it is able to find.
Any bug that requires as input a different or more complicated structure, such as redefining variables, will not be found. 
Furthermore, Fuzzilli will never create nested expressions and cannot output many JavaScript syntax errors.

Unsurprisingly, there are bugs that do require incorrect semantics or even incorrect syntax, as well as bugs that require unusual constructs.
We will briefly look at an example of such a bug that was found in V8.
Chromium issue 800032~\cite{lokibug} describes a high-impact bug found in V8 that could lead to remote code execution (RCE).
Note that although the bug has high impact with potential for RCE, no CVE was assigned to it, as it was discovered internally by Google Project Zero member Jung Hoon Lee.
The bug report includes the proof-of-concept in Listing~\ref{lst:loki_poc}, which triggers the issue.

\begin{figure}
  \centering
\begin{minted}{js}
class Sub extends RegExp {
  constructor(a) {
    // expected_nof_properties() skipped
    // due to error
    const a = 1; // semantic error
  }
}

let o = Reflect.construct(RegExp,[],Sub);
// OOB write
o.lastIndex = 0x1234;
\end{minted}
  \captionof{listing}{Proof-of-concept code for Chromium Issue 800032. This code triggers an error, which causes a miscalculation in the number of properties leading to an exploitable out-of-bounds write.}
  \label{lst:loki_poc}
\end{figure}

The proof-of-concept code creates a subclass of a Regular Expression object. In the constructor of the subclass, there is an error.
Specifically, the line \texttt{const a = 1} will attempt to redefine \texttt{a} as constant, which is invalid.
Because of this error, the size of an object gets incorrectly computed, which can then lead to out-of-bounds reads and writes on the object.
A fuzzing approach that follows a strict grammar definition would not be able to find issues such as this one.

\begin{figure}
  \centering
\begin{minted}{js}
function f() {
  ({
    a: {
      b = 0x1111, // invalid assignment
      c = 0x2222,
    }.c = 0x3333
  } = {});
}

f();
\end{minted}
  \captionof{listing}{Proof of concept code for CVE-2017-8729, which was caused by a parser error in Edge. Line 4 (\texttt{b = 0x1111}) contains a syntax error by trying to assign to a member with \texttt{=} while creating an object.}
  \label{lst:chakra_parse}
\end{figure}

Another example of a bug that could be difficult to find with a grammar-based fuzzer that strictly follows its grammar is shown in Listing~\ref{lst:chakra_parse}.
This example is CVE-2017-8729 of Edge~\cite{chakraparsebug}, where the parser would incorrectly parse the code, and, in doing so, lead to a type confusion when assigning to the object member later.
As this bug requires incorrect syntax to trigger, this example represents another case in which grammar-based fuzzers may suffer due to their adherence to the grammar.


We have just shown how grammar-based fuzzers may be unable to find certain bugs in interpreters, and previously, we showed how byte-level fuzzers, such as AFL, struggle to make any progress in fuzzing language-based inputs.
It is apparent there is a need for a new approach that can make progress and explore interpreters effectively, but without the limitations of a grammar.
In order to find a way to utilize the powerful evolutionary capabilities of tools like AFL on interpreters inputs, we introduce a new technique, called \emph{\methodname{}}.
\methodname{} works at a higher level than bytes, but is not strictly bound to the language grammar, allowing it to find bugs that neither byte-level fuzzing nor grammar-based fuzzing would find.

\section{Overview of Token-Level Fuzzing}
\label{sec:overview}

The idea behind \methodname{} is fairly simple: Valid tokens should be replaced with valid tokens.
For example, when fuzzing the code shown in Section~\ref{sec:motivation}, instead of mutating individual characters in the word \texttt{while}, we would replace the entire word with a different word.
If we replace \texttt{while} with \texttt{if} or \texttt{Number}, we would obtain a much better mutation.
Below are examples of possible better mutations if we use \methodname{}:

\begin{lstlisting}[escapeinside={(*}{*)}]
while (bar.x) (*$\longrightarrow$*) if (bar.x)
              (*$\longrightarrow$*) Number (bar.x)
              (*$\longrightarrow$*) while (bar+x)
              (*$\longrightarrow$*) while (while.x)       
\end{lstlisting}

Notice that \methodname{} can still produce invalid syntax, as is the case with the last line above: \texttt{while (while .x)}.
Even mutations like that can be beneficial if they trigger a new error handler or if they can iteratively be mutated until a different valid JavaScript statement is reached. 


A natural question to ask is how this technique compares to the ``dictionary'' that tools such as AFL~\cite{afldict} and LibFuzzer~\cite{libfuzzeroverview} allow users to provide.
The first major difference is that AFL will still perform the byte-level mutations as well as the dictionary-based mutations.
Second, the dictionary mutations are not aligned to tokens, so the fuzzer might insert the word \texttt{while} in the middle of a token instead of replacing the entire token.
Finally, it may take multiple token additions/replacements to reach a new and interesting input. 
Some fuzzers, such as AFL, will only insert one dictionary word in a mutation, which limits its exploration.

Another question is how \methodname{} compares to grammar-based fuzzing.
Grammar-based fuzzing mutates inputs or generates inputs according to a grammar, whereas \methodname{} does not follow any grammar.
\methodname{} can generate many patterns that can be difficult or impossible to produce for a particular grammar-based fuzzer, in particular those with complex or incorrect syntax.
On the other hand, grammar-based fuzzers focus on exercising the interpreter with correct syntax, possibly allowing faster exploration of that part of the program.
As a result, we expect that the two approaches complement each other well and are likely to find different bugs.

\begin{figure}
	\centering
	\caption{The architecture of \toolname{}. The tool has two primary components: The pre-parser and the fuzzing engine. The pre-parser is responsible for transforming input seeds into a list of 16-bit numbers. Then the fuzzing engine works on these lists, only decoding them back to JavaScript when they are passed to the target program.}
	\label{fig:architecture}
	\includegraphics[trim=0 0 0 0,clip,width=.9\linewidth]{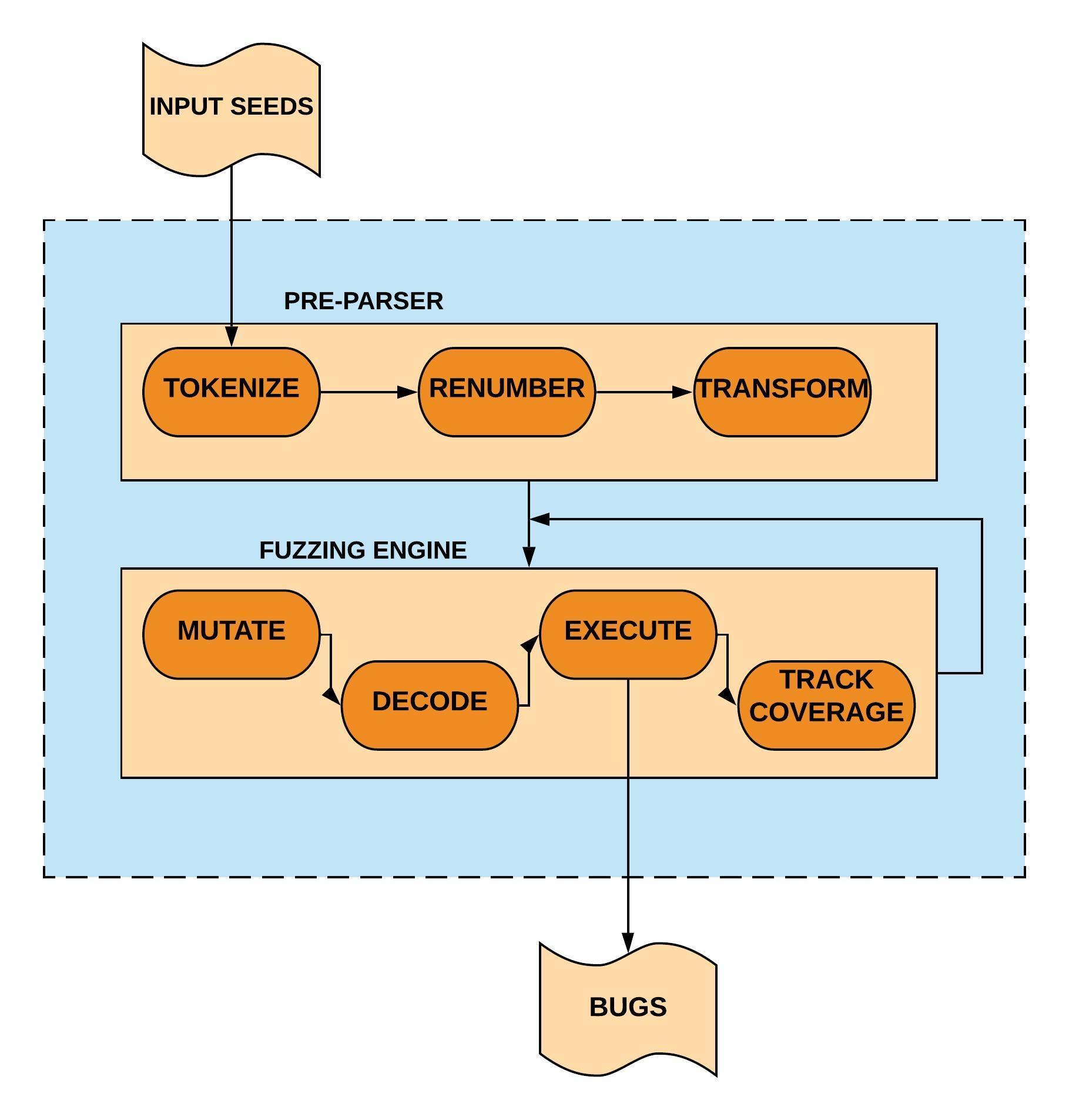}
\end{figure}

\begin{figure*}
	\centering
	\caption{An example of what happens to a single seed in \methodname{}. The seed first goes through the renaming and encoding stages which produce a list of numbers. Then when running in the fuzzing loop, it is mutated and decoded prior to execution, where coverage feedback will determine if the input is added to the queue or mutated further. We highlighted how changing a couple numbers in the encoded form results in completely different tokens in the decoded result.}
	\label{fig:overview}
	\includegraphics[trim=0 0 0 0,clip,width=\linewidth]{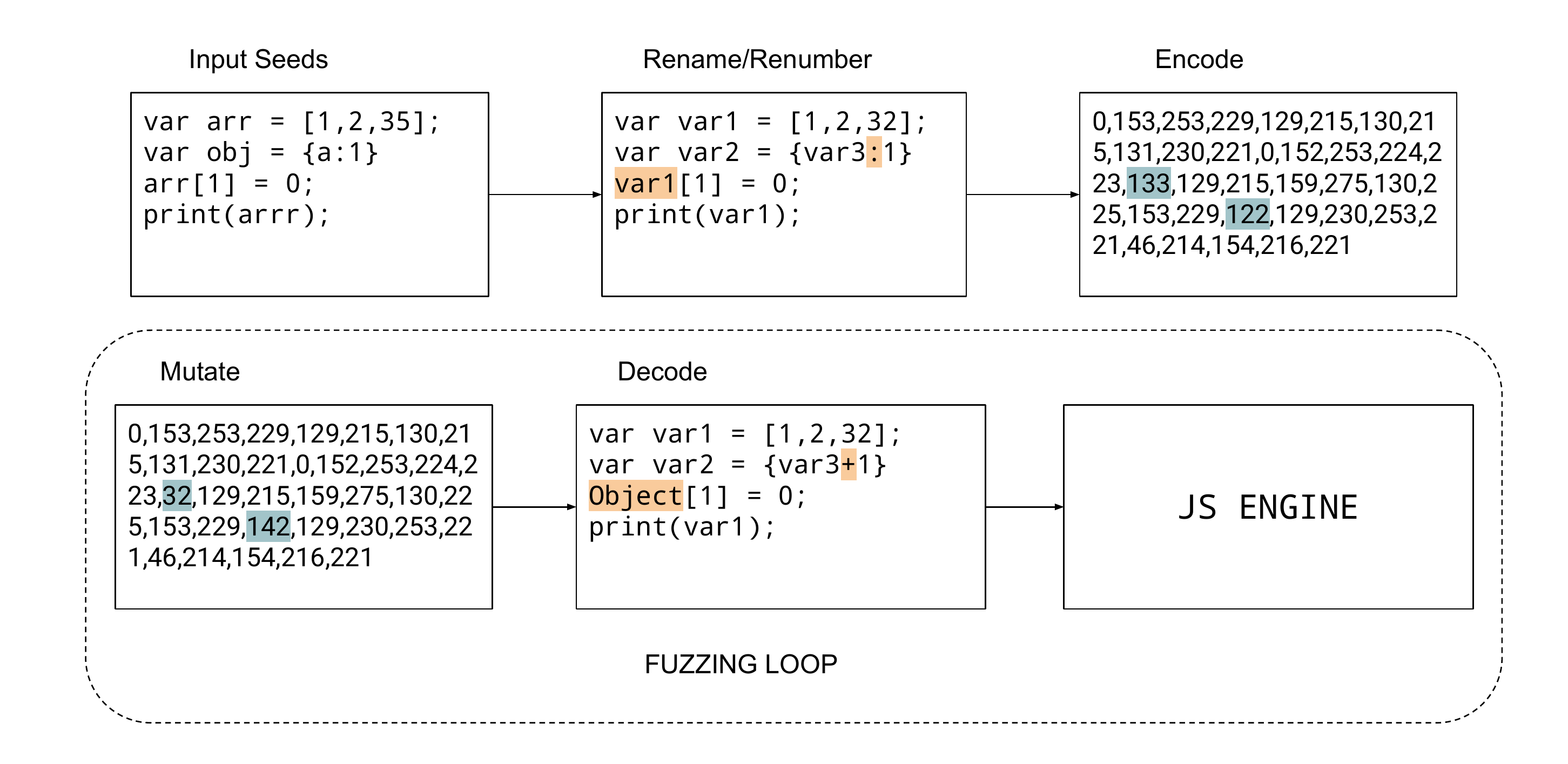}
\end{figure*}

\subsection{Method}
To create a fuzzer that works at the \emph{token level}, we start by constructing a map, which assigns to each possible token in the language a unique numerical value.
Then, we \emph{encode} input files into a list of numbers, which are the encoded version of the seeds.
Fuzzing is then performed on this list of numbers, where changing any number to a different number is equivalent to replacing the encoded token with a different token.
Whenever we want to run against the target (i.e., the JavaScript interpreter) we need to transform the mutated list of numbers back into the original language. 
This is done with a \emph{decode} function that replaces each number with the corresponding token and concatenates them with spaces as needed.
Thus, fuzzing can be done on the list of numbers without any knowledge of what they mean.

Of course, we need to consider that the list of valid tokens is infinite for many languages, as it includes, for example, all possible numbers and all possible variable names that are legal in the language.
If the token-map contains too many numbers, then it would unnecessarily slow down the fuzzer, because most tokens would be numbers and only very few would be other functionality.
To address this issue, we pick a small set of valid numbers consisting of all the powers of two (up to $2^{32}$), as well as the numbers that are a power of two plus/minus one. 
Similar values have been chosen for other fuzzers, such as DIFUZE and AFL, to reduce the number of possible inputs~\cite{corina2017difuze, afl}.
Similarly, we found by looking through regression cases that only a small number of variables were needed to trigger most bugs, so we limited the number of variable names to fifteen.

\subsection{Implementation}

Our implementation of \methodname{} is done on top of AFL, to take advantage of its coverage-guided engine.
The resulting tool is called \emph{\toolname{}}.
\toolname{} is the combination of two components: a preprocessor, written in Python, that analyzes the tokens of the input files and encodes them for fuzzing, and a modified version of AFL that performs fuzzing on the encoded inputs.
Figure~\ref{fig:architecture} shows the overall architecture of the tool. 

The preprocessor runs the following steps:

\begin{description}
    \item[Rename:] For each input seed, variable names are randomly replaced with one of the fifteen predefined variable names: var1, var2, ..., var15.
    Variable names are not repeated unless all fifteen variable names have been used.
    Within a single seed, all instances of the same variable are mapped to the same predefined name.
    \item[Renumber:] As described earlier, we limit the set of valid numbers to a predefined set.
    All numbers are replaced with the closest number from that set.
    \item[Token Analysis:] We use a JavaScript lexer to find all the tokens used in all the seeds. We assign to each token a numerical value, which will be its encoded value.
    \item[Encoding:] We transform each input into a list of numbers by replacing each token with its encoded value. This list is then flattened by encoding each value as a 16-bit integer.
\end{description}

After the preprocessing step, which generates the token mapping and the encoded seeds, fuzzing occurs on the encoded inputs.
This step is done with minor modifications to AFL:
\begin{description}
    \item[Mutations:] Mutations are slightly modified to work on an array of 16-bit numbers rather than an array of bytes. 
    16-bit numbers were necessary because there were more than 256 tokens. Note that this change is very small; it is effectively just changing the type of the array from \texttt{byte*} to \texttt{short*}.
    \item[Decoding:] The input is decoded immediately before executing the input in the target JavaScript interpreter. This small shim simply concatenates the tokens together, adding spaces as needed\footnote{No spaces are added for certain tokens, such as quotation marks.}.
\end{description}

Figure~\ref{fig:overview} shows an example of the various steps that \toolname{} applies to an input file. Although we expect that the input seeds are broad enough to include all valid tokens, if the seeds do not include all of the valid tokens, it is easy to add the remaining tokens by hand.

\subsection{Further Mutation Modifications}
Some of the mutations that AFL performs are not very useful or applicable to \methodname{}.
Specifically, these are the ``arithmetics'' and ``interesting number'' strategies.
According to these strategies, AFL will try inserting interesting numbers such as "1024", "2147483647", "-100663046", etc, into the stream of bytes.
Because these will get translated into a series of tokens, this just will add a constant random list of tokens into the fuzzed input.
Therefore, we removed these strategies from AFL, as they do not apply well to our scenario.

Of course, the next question is whether there are strategies that we can add to improve \methodname{}.
One simple strategy that we identified is to randomly insert and overwrite multiple tokens in a row.
The intuition behind this is that changing one token at a time may not be enough, and it may be necessary to change more than one token to create a new interesting input.
We tested with different numbers and found that inserting and overwriting up to three tokens at a time gave good results.
Therefore, we limit these to three tokens in our implementation.

We also wanted to improve AFL's ability to chain together different actions.
The reason for this is that there are many bugs in interpreters that require multiple actions chained together in the correct order.
Therefore, we added a mutation strategy that would copy a statement from one input to another.

We summarize the mutation strategies we added here:

\begin{description}
    \item[Random Insert:] Randomly insert new tokens somewhere into the file being mutated.
    \item[Random Overwrite:] Randomly overwrite tokens in a row in the file with the same number of new tokens.
    \item[Random Replace:] Randomly replace tokens in the file with new tokens. Note that this strategy can insert more or less tokens than were removed.
    \item[Statement Splice:] Copy a statement from one test case to another test case. This mutation strategy assumes that statements start and end at semicolons; the strategy then replaces all the tokens between two semicolons.
\end{description}

\section{Evaluation}
\label{sec:evaluation}

To evaluate our implementation of \methodname{}, we run the fuzzer on the JavaScript interpreters from the four major browsers, namely, V8, SpiderMonkey, JavaScriptCore, and ChakraCore\footnote{ChakraCore is no longer used in Edge as of January 2020~\cite{engadgetedget}.}.
Our goal is to understand the bug-finding capabilities of \methodname{} as well as how our implementation compares to other state-of-the-art JavaScript interpreter fuzzers. 
In order to reason about these goals, we answer the following research questions:
\begin{description}
\item[RQ1:] Does\methodname{} generate more syntactically correct inputs than byte-level fuzzing?
\item[RQ2:] How does \methodname{} compare to other state-of-the-art fuzzers?
\item[RQ3:] Is \methodname{} able to find real-world vulnerabilities in the latest JavaScript interpreters?
\item[RQ4:] Do bugs found by \methodname{} involve incorrect syntax/semantics?
\end{description}

\subsection{Experiment Setup}
\label{sec:experimentsetup}

We started by downloading the latest available versions of the four major JavaScript interpreters as of October 1, 2019.
These were the development versions cloned from the official git repositories.
We compiled all interpreters with debug checks.
Debug checks are additional checks that the programmers include to try to catch unexpected conditions~\cite{chromedcheck}; therefore, we enabled them to catch more potential security bugs.
We did not enable Address Sanitizer or other sanitizers, as these tended to be too slow in our tests.

\paragraph{Seed Collection.} Having good seeds is essential for our fuzzer, for multiple reasons. 
First, the list of potential tokens that will be used by our fuzzer are taken from the set of input files. 
Thus, it is essential that the seeds cover as many of the tokens used by the language as possible.
Second, our implementation of \methodname{} is based on AFL and evolutionary fuzzing, so having a quality set of diverse seeds helps the fuzzer greatly, because it will explore starting from these initial seeds.
To collect seeds, we manually selected regression tests from the repositories of the various JavaScript interpreters.
We manually picked seeds covering a wide range of functionality, but limited the number of seeds to one hundred.

\paragraph{Comparison with other tools.} We compared \toolname{} against the following state-of-the-art tools: AFL~\cite{afl}, Fuzzilli~\cite{gross2018fuzzil}, CodeAlchemist~\cite{han2019codealchemist}, and Superion~\cite{wang2019superion}. 
To this end, we ran each tool for three days on 30 cores, on each of the four major JavaScript interpreters, resulting in a total of 2,160 core-hours for each fuzzing run.
We then repeated each fuzzing run (that is, each fuzzer-JavaScript interpreter combination) five times, to limit randomness in our experiments.
Note that Fuzzilli does not provide a mechanism for using seeds, so it was run without seeds.
On the other hand, the authors of CodeAlchemist used far more seeds in their paper~\cite{han2019codealchemist}, and, therefore, to fairly evaluate this tool, we created a much larger seed collection, which included all JavaScript files from the regression tests, resulting in 32,682 seeds. 
This larger set of seeds was only used when testing CodeAlchemist.

Note that when comparing against these tools there may be a bias in terms of number of bugs found.
This is because other published tools may have already reported the bugs that they were able to find, and these bugs might have been already fixed in the JavaScript engines that we analyzed.
However, running experiments on the latest JavaScript interpreters will let us know if \toolname{} finds different bugs than the other tools.

\subsection{Syntactically Valid Inputs}

The most basic assumption of \methodname{} is that it generates more syntactically correct inputs than byte-level fuzzing, and that these inputs will, in turn, trigger deeper functionality.
To assess the validity of this assumption, we first compare the results of AFL and \toolname.
Both fuzzers were given the same seeds, and AFL was given all of the tokens in the input files as a dictionary.
With a dictionary, AFL will try inserting the keywords in the mutation steps.
This allows AFL to make some progress on languages such as JavaScript, and showcases the best configuration for AFL~\cite{afldict}.
In our experiments, even with a full dictionary and the same input seeds, AFL was only able to find 2 bugs across all the JavaScript interpreters, whereas \toolname{} found 19.
Furthermore, as shown in Table~\ref{tab:Bug Breakdown}, both bugs reported by AFL were also found by our tool.

Next, we added tracking to the V8 JavaScript engine to determine how many of the inputs, generated by the two fuzzers, were parsed successfully or led to parser errors. 
These numbers are shown in Table~\ref{tab:errorrates}.
Only 10.7\% of all the inputs AFL produced could be parsed successfully.
This shows that, as we suspected in Section~\ref{sec:motivation}, most inputs generated by AFL fail to parse, and do not trigger any reasonable functionality in the JavaScript interpreters.
The improvement provided by \methodname{} is immediately evident; 29.98\% of all inputs generated by \toolname{} were successfully parsed. 
The higher fraction of successfully parsed inputs allows the fuzzer to generate more inputs that trigger useful functionality.
This, in turn, allows the fuzzer to find deeper bugs and explore more of the JavaScript interpreter's functionality.

\begin{mdframed}
  \textbf{Answer for RQ1}: The results show that \methodname{} generates syntactically correct inputs about three times more often than byte-level fuzzing, enabling more efficient fuzzing of interpreters.
\end{mdframed}

\begin{table}

\caption{
This table shows what fraction of inputs generated by AFL and \toolname{} are able to be parsed successfully when fuzzing V8. 
The higher parse rate of \toolname{} shows that by mutating tokens instead of bytes, our technique is able to generate more correct inputs.}
\label{tab:errorrates}

\begin{tabular}{l|l}
\textbf{Fuzzer}                & \textbf{Successful Parse Rates} \\ \hline
\textbf{AFL}             & 10.70\%  \\
\textbf{Token-Level AFL} & 29.98\%  \\ \hline
\end{tabular}
\end{table}

\subsection{Comparison with other State-of-the-Art Fuzzers}
\label{sec:comparison}
In this section, we will explore how \toolname{} performs when compared against other state-of-the-art JavaScript interpreter fuzzers.
For this comparison, we selected AFL, Fuzzilli~\cite{gross2018fuzzil}, CodeAlchemist~\cite{han2019codealchemist}, and Superion~\cite{wang2019superion}.
These comparison tools were chosen because of their impressive results and their varying techniques.

As mentioned previously, we evaluated all of these fuzzers on the latest available JavaScript interpreters, which were retrieved from the official repositories on Oct 1, 2019. Each fuzzer was run on 30 cores for three days on each of the four JavaScript interpreters. Moreover, each test run was repeated five times to reduce randomness.

As is usual for fuzzing research, we use the number of bugs found as the main performance metric.
For the purpose of this analysis, we consider any debug check, release check, or memory corruption to be a bug. 
Although debug checks may not always indicate that a security issue was found, they do indicate that an assumption was violated, and they show that a fuzzer is finding bugs that have not been previously found.
To identify unique bugs, we filter the tool's reports based on any asserts hit, as well as using manual analysis to ensure that only unique issues are counted.

Additionally, we investigate block coverage during this evaluation.
Although block coverage may not be as meaningful a measurement as the number of bugs found, it still shows useful information~\cite{klees2018evaluating, salls2020exploring}. 
To be able to trigger a bug, a fuzzer must be able to reach the code where the bug is located.
So, coverage is a necessary, but not sufficient, condition for finding bugs and can be used as a performance metric.
We collected block coverage information throughout each of the fuzzing runs.

\begin{figure}[t]
	\centering
	\caption{This graph shows the total number of unique bugs found by each of the tested fuzzers when run on the four major JavaScript interpreters for a time period of 72 hours. This graph shows the aggregate number of bugs across all five runs, and only unique bugs are counted.}
	\label{fig:bugscomparison}
	\includegraphics[trim=0mm 0 0 0,clip,width=\linewidth]{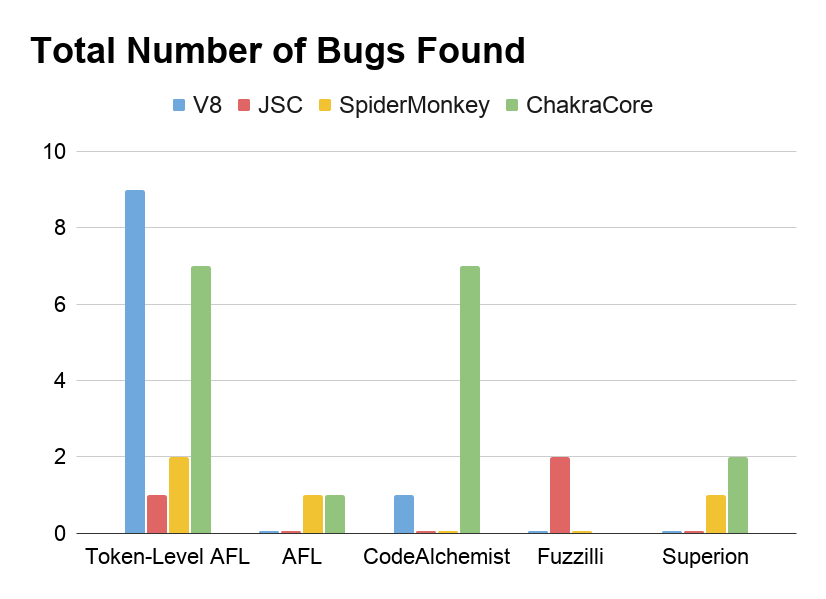}
\end{figure}

\textbf{Results:} As shown in Figure~\ref{fig:bugscomparison}, \toolname{} found the most crashes during the three-day fuzzing periods. 
More precisely, \toolname{} found 19 total bugs across the five runs, while the second best performer, CodeAlchemist, found 8 bugs.
Additionally, only 4 of the 19 bugs found by \toolname{} were found by any other tool (see Table~\ref{tab:Bug Breakdown}.)
Each of the other 15 bugs were unique to \toolname.
Also, although CodeAlchemist found seven bugs in ChakraCore, only two of those bugs overlapped with the four found by \toolname{}.
This indicates that our method finds bugs that other fuzzers are not able to find.
Furthermore, in Table~\ref{tab:averagecrashes} we show the average number of bugs each tool found in any single run.
This data shows that \toolname{} also finds more bugs in each run than other tools.

\begin{table}[t]
\small
\caption{
	 Average number of bugs found by each of the tested fuzzers on the four major JavaScript interpreters in a single run. The 95\% confidence interval is $\pm1.39$ for each entry. (Fuzzilli does not have code for running on ChakraCore, so that table entry is omitted).}
\label{tab:averagecrashes}

\begin{tabular}{l|lllll}
                         & V8   & JSC  &  \begin{tabular}[c]{@{}l@{}}Spider- \\ Monkey  \end{tabular} & \begin{tabular}[c]{@{}l@{}}Chakra\end{tabular} \\ \hline
\textbf{Token-Level AFL} & 5.2  & 0.6 & 0.8           & 3.2          \\
AFL                      & 0    & 0    & 1.0          & 0.2       \\
CodeAlchemist            & 0.6  & 0    & 0            & 4.0       \\
Fuzzilli                 & 0    & 1.0  & 0            & N/A       \\
Superion                 & 0    & 0  &  1             & 0.4       \\
\end{tabular}
\end{table}

\begin{figure}[t]
	\centering
	\caption{This graph shows the block coverage over time for each of the fuzzers when running on V8. \toolname{} was able to continually find and trigger new blocks throughout the three-day experiment.}
	\label{fig:v8coverage}
	\includegraphics[trim=0mm 0 0 0,clip,width=\linewidth]{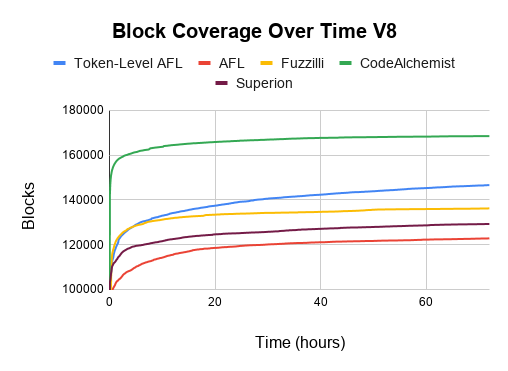}
\end{figure}

\begin{table}[t]
\small
\caption{
	Average number of basic blocks triggered by each of the tools on each of the target interpreters. \toolname{} performed similarly to Fuzzilli in terms of number of blocks covered. CodeAlchemist, which used many more seeds, had the best block coverage.}
\label{tab:coverage}

\begin{tabular}{l|llll}
                & V8     & JSC    & \begin{tabular}[c]{@{}l@{}}Spider- \\ Monkey  \end{tabular} & \begin{tabular}[c]{@{}l@{}}Chakra- \\ Core\end{tabular} \\ \hline

\textbf{Token-Level AFL} & 146,625 & 246,720 & 172,050 & 178,126     \\
AFL                      & 122,833 & 219,774 & 157,031 & 139,070  \\
CodeAlchemist            & 168,512 & 256,650 & 212,267 & 214,499    \\
Fuzzilli                 & 136,218 & 244,391 & 184,363 & N/A       \\
Superion                 & 129,753 & 223,656 & 165,674 & 169,440
\end{tabular}
\end{table}

\begin{table}[t]
\footnotesize
\setlength{\tabcolsep}{4pt}
\caption{
This table shows a breakdown of which fuzzers found each of the 27 unique bugs during the comparison experiment on the October 2019 versions. This shows cumulative results across all 5 runs.}
\label{tab:Bug Breakdown}
\begin{tabular}{clccccc}
\multicolumn{1}{l}{}                                                       &        & \textbf{\begin{tabular}[c]{@{}c@{}}Token-\\ Level\\ AFL\end{tabular}} & AFL                       & \begin{tabular}[c]{@{}c@{}}Code-\\ Alchemist\end{tabular} & Fuzzilli                  & \begin{tabular}[c]{@{}c@{}}Superion\end{tabular} \\ \hline
                                                                           & Bug 1  & \cellcolor[HTML]{C9DAF8}X                                             & \multicolumn{1}{l}{}      & \multicolumn{1}{l}{}                                          & \multicolumn{1}{l}{}      & \multicolumn{1}{l}{}                                 \\
                                                                           & Bug 2  & \cellcolor[HTML]{C9DAF8}X                                             &                           &                                                               &                           &                                                      \\
                                                                           & Bug 3  & \cellcolor[HTML]{C9DAF8}X                                             &                           & \cellcolor[HTML]{F4CCCC}X                                     &                           &                                                      \\
                                                                           & Bug 4  & \cellcolor[HTML]{C9DAF8}X                                             &                           &                                                               &                           &                                                      \\
                                                                           & Bug 5  & \cellcolor[HTML]{C9DAF8}X                                             &                           &                                                               &                           &                                                      \\
                                                                           & Bug 6  & \cellcolor[HTML]{C9DAF8}X                                             &                           &                                                               &                           &                                                      \\
                                                                           & Bug 7  & \cellcolor[HTML]{C9DAF8}X                                             &                           &                                                               &                           &                                                      \\
                                                                           & Bug 8  & \cellcolor[HTML]{C9DAF8}X                                             &                           &                                                               &                           &                                                      \\
\multirow{-9}{*}{V8}                                                       & Bug 9  & \cellcolor[HTML]{C9DAF8}X                                             & \multicolumn{1}{l}{}      & \multicolumn{1}{l}{}                                          & \multicolumn{1}{l}{}      & \multicolumn{1}{l}{}                                 \\ \hline
                                                                           & Bug 1  & \cellcolor[HTML]{C9DAF8}X                                             & \cellcolor[HTML]{D9EAD3}X & \multicolumn{1}{l}{}                                          & \multicolumn{1}{l}{}      & \cellcolor[HTML]{D9D2E9}X                            \\
\multirow{-2}{*}{\begin{tabular}[c]{@{}c@{}}Spider-\\ monkey\end{tabular}} & Bug 2  & \cellcolor[HTML]{C9DAF8}X                                             & \multicolumn{1}{l}{}      & \multicolumn{1}{l}{}                                          & \multicolumn{1}{l}{}      & \multicolumn{1}{l}{}                                 \\ \hline
                                                                           & Bug 1  &                                                                       &                           &                                                               & \cellcolor[HTML]{FFF2CC}X &                                                      \\
                                                                           & Bug 2  & \cellcolor[HTML]{C9DAF8}X                                             &                           &                                                               &                           &                                                      \\
\multirow{-3}{*}{JSC}                                                      & Bug 3  &                                                                       &                           &                                                               & \cellcolor[HTML]{FFF2CC}X &                                                      \\ \hline
                                                                           & Bug 1  & \cellcolor[HTML]{C9DAF8}X                                             & \cellcolor[HTML]{D9EAD3}X & \cellcolor[HTML]{F4CCCC}X                                     & \multicolumn{1}{l}{}      & \cellcolor[HTML]{D9D2E9}X                            \\
                                                                           & Bug 2  & \cellcolor[HTML]{C9DAF8}X                                             &                           & \cellcolor[HTML]{F4CCCC}X                                     &                           & \multicolumn{1}{l}{}                                 \\
                                                                           & Bug 3  & \multicolumn{1}{l}{}                                                  &                           & \multicolumn{1}{l}{}                                          &                           & \cellcolor[HTML]{D9D2E9}X                            \\
                                                                           & Bug 4  &                                                                       &                           & \cellcolor[HTML]{F4CCCC}X                                     &                           &                                                      \\
                                                                           & Bug 5  &                                                                       &                           & \cellcolor[HTML]{F4CCCC}X                                     &                           &                                                      \\
                                                                           & Bug 6  &                                                                       &                           & \cellcolor[HTML]{F4CCCC}X                                     &                           &                                                      \\
                                                                           & Bug 7  &                                                                       &                           & \cellcolor[HTML]{F4CCCC}X                                     &                           &                                                      \\
                                                                           & Bug 8  &                                                                       &                           & \cellcolor[HTML]{F4CCCC}X                                     &                           &                                                      \\
                                                                           & Bug 9  & \cellcolor[HTML]{C9DAF8}X                                             &                           &                                                               &                           &                                                      \\
                                                                           & Bug 10 & \cellcolor[HTML]{C9DAF8}X                                             &                           &                                                               &                           &                                                      \\
                                                                           & Bug 11 & \cellcolor[HTML]{C9DAF8}X                                             &                           &                                                               &                           &                                                      \\
                                                                           & Bug 12 & \cellcolor[HTML]{C9DAF8}X                                             &                           &                                                               &                           &                                                      \\
\multirow{-13}{*}{Chakra}                                                  & Bug 13 & \cellcolor[HTML]{C9DAF8}X                                             & \multicolumn{1}{l}{}      & \multicolumn{1}{l}{}                                          & \multicolumn{1}{l}{}      & \multicolumn{1}{l}{}                                 \\ \hline
\end{tabular}
\end{table}

When investigating coverage, we found that, on average, \toolname{} covered more blocks in the JavaScript interpreters than three of the other tools; only CodeAlchemist triggered more basic blocks.
The average number of basic blocks found in each configuration is shown in Table~\ref{tab:coverage}, and a graph of block coverage over the three days of fuzzing is shown in Figure~\ref{fig:v8coverage}.
When investigating these numbers in more detail, we discovered that the seeds may play a large role in CodeAlchemist's superior code coverage.
In particular, the 32,682 seeds given to CodeAlchemist alone triggered about 160,000 blocks in V8, whereas the 100 seeds given to \toolname{} only covered about 94,000 blocks. 
However, even with higher coverage, CodeAlchemist triggered fewer bugs, showing that code coverage does not yield bugs on its own; assumptions must be violated as well.
Also, \toolname{} was able to find many blocks that were not triggered by the initial seeds.
Finally, the graph shows that \toolname{} was still finding new basic blocks at the end of the fuzzing period, whereas the other tools had plateaued.

It is worth mentioning that the lack of bugs found by other tools does not necessarily indicate a lack of performance.
Instead, it is quite likely that, because these fuzzers are open-source, they are currently being run and bugs that they find are reported and fixed frequently.
However, our results do show that \toolname{} is finding new, different bugs that these other tools are not able to find as easily.

\textbf{Breakdown:} Table~\ref{tab:Bug Breakdown} shows the overlap of bugs found during the comparison experiments.
In V8 and Spidermonkey, \toolname{} was the only fuzzer to find unique bugs that no other fuzzer found.
However, in both JSC and Chakra there were some bugs that we missed and that were found only by a different fuzzer.

We performed deeper analysis to understand if there are similarities among the bugs that only \toolname{} finds, as well as those bugs that our system missed. 
For many of \toolname{}’s unique crashes, we found that the inputs used to trigger these bugs include code patterns with uncommon or completely invalid syntax. 
This underlines the value of a fuzzer that can generate inputs that do not strictly follow a grammar. 
We provide a more detailed case study for some of these crashes in Section~\ref{case_study}. 

For the bugs that \toolname{} did not find, we did not find any obvious shared characteristics. 
In fact, it appeared to us that it would be possible to trigger these crashes with different seeds or better luck. 
The bugs were triggered by specific sequences of (valid) operations, and \toolname{} had not (yet) produced the required order.

\begin{mdframed}
  \textbf{Answer for RQ2}: \toolname{} is able to find bugs that other state-of-the-art fuzzers are unable to find. 
  Furthermore, in our tests, \toolname{} found more bugs in the major JavaScript interpreters than any of the other state-of-the-art fuzzers.
\end{mdframed}

\begin{table*}
\caption{
	This table shows the bugs which \toolname{} found in the analyzed JavaScript interpreters over a 60-day period. 
	Some of these bugs resulted in memory corruption, which could lead to exploitation and remote code execution.
	In the ``Status'' column we note if we have confirmed that the bug still exists in the most up-to-date code, reported it, or if it was fixed internally. 
	We are currently in the process of responsibly disclosing all confirmed bugs to the respective software vendors.}
\label{tab:realbugs}
\small

\begin{tabular}{lllll}
\hline
\textbf{Bug Number} & \textbf{JS Interpreter} & \textbf{Description}                                                                & \textbf{Status} & \textbf{Bug ID} \\ \hline
1                  & V8                 & Memory corruption while parsing             & Reported/Fixed  & CR 1015567          \\
2                  & V8                 & Debug Check due to incorrect parsing of arrow functions.                            & Reported/Fixed  & V8 9758             \\
3                  & V8                 & Null dereference      & Fixed Internally & \\
4                  & V8                 & Debug Check in regular expression runtime                                           & Reported/Fixed  & CR 1018592          \\
5                  & V8                 & Out of bounds indexing in an array due to incorrect parsing       & Reported/Fixed  & CR 1021457          \\
6                  & V8                 & Parser debug check due to incorrectly allocated variable       & Fixed Internally & \\
7                  & V8                 & Debug Check in garbage collection                                                   & Reported        & CR 1044261          \\
8                  & V8                 & Debug check when converting integer to index                                    & Fixed Internally        &            \\
9                  & V8                 & Triggers unreachable code due to frozen elements                                    & Reported/Fixed  & CR 1045572          \\
10                  & V8                 & Unexpected error handler triggered in JIT       & Fixed Internally & \\
11                 & V8                 & Check failed due to incorrect object size                                           & Reported/Fixed  & CR 1076106          \\
12                 & V8                 & JIT bug leading to memory corruption                                                & Reported/Fixed        & 
CVE-2020-6468   \\
13                 & V8                 & Triggers unreachable code due to frozen elements                                    & Reported/Fixed        & V8 10484            \\

14                 & V8                 & Jit bug in bytecode analysis                                    & Fixed internally        &            \\
15                 & V8                 & Parser error leading to debug check                                    & Confirmed in latest        &            \\
16                 & V8                 & JIT assertion related to a syntax error                                    & Confirmed in latest        &            \\

17                 & JSC                 & JIT bug resulting in an unexpected switch case                                    & Reported        & Webkit 221069          \\
18                 & JSC                 & JIT bug in FTL resulting in an unexpected null pointer                                    & Fixed internally        &            \\
19                 & JSC                 & JIT bug in DFG failing a validation check                                    & Confirmed in latest        &            \\
20                 & JSC                 & JIT bug in FTL to DFG Lowering
& Fixed Internally        &            \\
21                 & SpiderMonkey                 & Length related assertion
& Reported        &  1669616           \\
22                 & SpiderMonkey                 & Parser assertion
& Confirmed in latest        &            \\
23                 & SpiderMonkey                 & Parser bug leading to leaked magic value
& Fixed        &            \\
24                 & ChakraCore                 & Type mismatch in parsing
& Reported       &  MS 041681          \\
25                 & ChakraCore                 & Unexpected case in the JIT
& Reported        &  MS 041671         \\
26                 & ChakraCore                 & Array length changed where it should not have changed
& Reported        &  MS 041673          \\
27                 & ChakraCore                 & Out of Bounds in Array runction
& Reported        &  MS 041679          \\
28                 & ChakraCore                 & Assertion setting a field on an object 
& Reported        &  MS 041678          \\
29                 & ChakraCore                 & Assertion in set accessor 
& Reported        &  MS 041676          \\
\hline
\end{tabular}
\end{table*}

\subsection{Real-World Bugs}

In the previous section, we have shown that \toolname{} is effective in finding bugs in JavaScript interpreters that other fuzzers are unable to find.
These bugs were in the JavaScript interpreters that were available as of October 1, 2019.
In a separate experiment, we wanted to further explore \toolname's ability to find bugs when run over a longer period of time (rather than the three days used for the comparative evaluation). To do this, we let our fuzzer run for 60 days. We started with the interpreters as of September 20, 2019. Over the duration of the following two months, we periodically restarted the fuzzer and updated the JavaScript interpreters as new versions became available.

Table~\ref{tab:realbugs} shows a summary of all the bugs that \toolname{} found across the analyzed JavaScript interpreters.
The table shows in which interpreter each bug was found and a description of the bug.
The status column shows if the bug has been reported by us and fixed. 
``Confirmed'' indicates that we have confirmed the bug in the latest version.
``Fixed internally'' means that the interpreter developers identified and fixed the bug without our report (i.e., after we found the bug, but before we had a chance to report it); sometimes these were short-lived bugs.
Note that the 19 bugs found during our comparative evaluation by \toolname{} were also found during this experiment, and they are included in Table~\ref{tab:realbugs}. Thus, \toolname{} identified 10 additional bugs when given more time.

Our fuzzer found the \numbugs{} bugs across many areas of the JavaScript interpreters: from the parser, to the handler of regular expressions, to the JIT compiler.
We believe that this shows not only that \toolname{} is capable of finding unknown bugs in JavaScript interpreters, but also that it is widely applicable and can find bugs in many components of the interpreter.

Also, these bugs include some that could lead to remote code execution.
We were able to write an RCE exploit for Chrome using bugs that we found with this tool.
Furthermore, we have been awarded over ten thousand dollars in bounties, showing the impact of our research.

\begin{mdframed}
  \textbf{Answer for RQ3}: \toolname{} is able to find real-world bugs in all of the major JavaScript interpreters. 
  This shows that \toolname{} has impact and can be used for finding previously unknown bugs as well as for catching bugs as they are introduced.
\end{mdframed}

\subsection{Case Study}
\label{case_study}

In this section, we investigate some of the bugs to determine if \toolname{} is finding bugs that involve invalid syntax, which strict grammar based tools may be unable to find. 

In Listing~\ref{lst:array_index}, we show (a minimized) example of the JavaScript code that triggers a bug that \toolname{} found.
This is a bug in V8 that leads to memory corruption. It
requires a syntax error to trigger and was introduced when new parser code was added that allowed for certain incorrect syntax patterns, such as the one that is shown in the listing.
Our tool was able to find this syntax, partially because of its evolutionary behavior.
The bug was fixed due to our report and a bounty was awarded.

\begin{listing}[tbp]
	\vspace{1.0mm}
\begin{minted}{js}
class var6 extends Object {
  constructor ( a,b,c) {
    super (1.1 ) 1 ;
  }
};

new var6();
\end{minted}
	\caption{Code which triggers a bug found by \toolname{} in V8. This bug contains a syntax error due to the number 1 after the call to \texttt{super(1.1)}. In this case, the parser would incorrectly calculate the index into an array, resulting in exploitable memory corruption.}
\label{lst:array_index}
\end{listing}

Listing~\ref{lst:gc} shows another example of code that triggers a bug. 
This bug results in a Debug Check in V8's garbage collector.
The code shown is a minimized version of the real test case, after removing redundant statements.
This bug is more complex than the previous example, and requires many valid JavaScript statements.
We attribute the ability of \toolname{} to produce complex valid test cases to its coverage-guided capabilities, which will tend to discard test cases that do not hit new functionality, allowing it to explore deep code paths.

\begin{listing}[tbp]
	\vspace{1.0mm}
\begin{minted}{js}
function f () { 
  var14=[1,2,3,4,5,6,7,8]; 
  var15=var14;
  var14.length = 0x100 ; 
  var14.__defineGetter__(/./, function(){
    var14.unshift ( 0x20 ) ;
    var14.shift(); 
    var var3=new Uint32Array(var14); 
    Object.entries(var14).toString(); 
  } ) ; 
  print(Object.entries(var14).toString());
} 
f();
\end{minted}
	\caption{This minimized test case triggers a debug check found in V8. This bug is caused by repeated shifting and unshifting of an array, which can trigger a debug check in the garbage collector.}
\label{lst:gc}
\end{listing}

Bugs found by our technique included both examples where incorrect syntax or semantics is used to trigger a bug and examples where no such error exists in the test case.
Also, many of the bugs that we found were in the parser, as opposed to the other tools we tested, which tend to miss those bugs.
Our results show that \toolname{} is applicable to finding bugs both in the parser and elsewhere in the JavaScript interpreter.

\begin{mdframed}
  \textbf{Answer for RQ4}: Bugs found by \toolname{} include examples where both entirely valid syntax is used and examples where invalid syntax is needed. 
\end{mdframed}





\section{Discussion}
\label{sec:discussion}

\methodname{} is a promising new technique that enables deep fuzzing of JavaScript interpreters, without some of the limitations that come with grammar-based fuzzers.
By performing coverage-guided mutations on tokens, rather than individual bytes, it can easily mutate the highly structured inputs involved in the language.
Additionally, because \methodname{} is able to find bugs with unusual constructs (syntax and semantics), we believe it will complement the current grammar-based approaches nicely.
In this section, we will discuss the generalizability of our technique, as well as directions for future work.

\subsection{Generalizability}

Although we implemented and tested \toolname{} only on JavaScript interpreters, the technique is likely applicable to other programs that process inputs formatted in well-defined languages, such as compilers and configuration parsers.
The tool would need a different pre-processor, specific to the target, that can separate the text into tokens and identify variables.
Similarly, a new decoder would need to be written for that target to transform the encoded input back into the original language.
These are not technical challenges, and we believe this technique should be effective on other token-based programs, especially given the results it has shown on JavaScript interpreters.
Furthermore, this is likely easier than adapting a grammar-based fuzzer to a new target.

\subsection{Seed Selection}

\toolname{} relies heavily on the input seeds, and it is intuitive that this selection can matter greatly.
If a seed is close to triggering a bug, then the number of mutations needed to exercise the bug may be small.
In fact, we noticed substantial similarities between some of the bugs that we found and the input test cases that we provided.
Additionally, having seeds that trigger a wide variety of functionality helps the fuzzer to explore the various areas of the interpreter's code.

One result shown in Section~\ref{sec:evaluation} is that \toolname{}'s block coverage could likely be improved by having a better, larger set of seeds.
For our experiments, we used a fairly ad hoc approach for our seed collection, and applying a better and more systematic methods might yield even better results.
For example, Skyfire~\cite{wang2017skyfire} could be used to generate promising JavaScript seeds.
There are also various papers suggesting better seed selection strategies, which we could employ to improve our results~\cite{Rebert:2014:OSS:2671225.2671280, cheng2019optimizing}.

\subsection{Future Work}

Because our technique transforms the JavaScript tokens into the familiar binary-based format, we could leverage recent advancements that have been made in the fuzzing field.
For example, because there are so many edges in the JavaScript interpreters, we find that there are many collisions in the edge tracking of AFL. 
We could use the path sensitivity of CollAFL~\cite{gan2018collafl} to help remedy this.
Applying ensemble based fuzzing~\cite{chen2019enfuzz}, by using \toolname{} alongside a grammar-based approach, could allow both techniques to build on top of their results.
Another direction would be to try to use better prioritization on the inputs, as suggested by Wang, et al.~\cite{wangnot}, especially since we typically have tens of thousands of inputs in the fuzzer queue after a few days of fuzzing.

\section{Conclusion}
\label{sec:conclusion}

In this paper, we have presented \methodname{}, a new technique for fuzzing language-based programs, such as interpreters.
\methodname{} allows one to fuzz these complex programs without the need of a grammar, allowing it to exercise both the parsing layers, as well as the actual interpretation.
This relatively simple idea (one can fuzz at an intermediate level between grammar-based and byte-based fuzzers) provides security researchers with a powerful new technique that can be built upon for further research.

In our evaluation, \toolname{} found \numbugs{} new bugs across the most up-to-date JavaScript interpreters, several of which were  high-severity issues.
Given the difficulty of fuzzing such programs, we believe that these results showcase the potential of our technique.

\bibliographystyle{IEEEtranS}
\bibliography{biblio}


\end{document}